\documentclass[11pt, conference, compsocconf]{IEEEtran}
\usepackage{epsfig,endnotes}
\usepackage{float}
\usepackage{subcaption}
\usepackage{algorithm}
\usepackage{algorithmic}
\usepackage{url}
\usepackage{balance}
\usepackage{amsmath}

\usepackage{balance}

\begin{document}



\title{Oseba: Optimization for Selective Bulk Analysis in Big Data Processing}

\author{Rui Wang and Jun Wang\\
Department of Electrical Engineering and Computer Science\\
University of Central Florida, Orlando, FL\\
ruiwang@knights.ucf.edu, jwang@eecs.ucf.edu
}

\maketitle

\begin{abstract}

Selective bulk analyses, such as statistical learning on temporal/spatial data, are fundamental to a wide range of contemporary data analysis. 
However, with the increasingly larger data-sets, such as weather data and marketing transactions, the data organization/access becomes more challenging in selective bulk data processing with the use of current big data processing frameworks such as Spark or keyvalue stores.         
In this paper, we propose a method to optimize selective bulk analysis in big data processing and referred to as Oseba. 
Oseba maintains a super index for the data organization in memory to support fast lookup through targeting the data involved with each selective analysis program.
Oseba is able to save memory as well as computation in comparison to the default data processing frameworks.


\end{abstract}



\section{Introduction}


In today's big data era, statistical/machine learning methods~\cite{price2000comparison, cite:PatternR, yin_time} plays an important role in many analysis activities, such as performance evaluation, knowledge discovery, sequential reasoning and prediction. 
For example, statistical  methods like Centered Moving Average or Stationarity Computation could be applied to investigate how the data changes within a period of time. 
Knowledge discovery methods such as trends/seasonality analysis, pattern extraction or distance comparison are usually applied to extract knowledge, which could be used to predict the future trend, e.g, weather forecast or stock price prediction. 
Moreover, machine learning algorithms like modeling training  usually group data into different parts in order to capture a precise prediction model.
With these analysis activities, overall or subset of data sets are often involved. And in this paper, we refer them as selective bulk data analysis.
However, with the increasingly larger data-sets~\cite{Wang2014}, 
the data organization/access becomes more challenging. 


Currently, MapReduce~\cite{dean2008mapreduce} is the de-facto programming model for big data processing and Spark is a popular open-source framework with the 
implementation of MapReduce. 
Spark provides Resilient Distributed Datasets (RDDs), which are partitioned in memory collections of data items, and allows the partitioned data to reside in memory for repeatedly or interactively processing. 
Spark provides an interface based on coarse-grained transformations to apply the same operation to all data partitions. 

Unfortunately, such a coarse-grained data processing fashion is inefficient in dealing with selective bulk data analysis.   
This is because, instead of applying operations to all data partitions, a selective bulk analysis program may only need to access part of the data throughout its entire execution. 
For instance, periods analysis such as Distance Comparison will only need the data in two specific periods. Thus, to prepare the data for selective bulk analysis programs, a large amount of computation and memory will be required to generate and store the corresponding involved data. 
One main challenge involved in selective bulk analysis is to find a method such that the required data partitions can be efficiently targeted and accessed.

In this paper, we implement Spark as the basic building block for our platform. This is because selective bulk analysis usually involves interactive analysis and data sets need to be accessed for multiple analysis on different partitions. It should be much more efficient when the data are resident in memory like Spark.
To optimize selective bulk analysis within Spark, we propose a method based on index for the in-memory partitioned data. This method enables us to save memory as well as computation as compared to the default data processing method.

\section{ Selective Bulk Analysis and In-Memory Data Processing}\label{background}


\textbf{Selective Bulk Analysis} 
In practice, given a temporal/spatial dataset such as time series, multiple interactive data analyses could be  involved. Specifically, we illustrate several common used methods in selective bulk analysis as following.   
\begin{itemize}
%
\item{\emph{Moving Average}} is often implemented in the analysis of a time series to smooth out short-term fluctuations and highlight longer-term trends or cycles. For example, a 10-day MA would average out the closing prices of a stock for the first 10 days as the first data point. The next data point would drop the earliest price, add the price on day 11 and take the average, and so on. 
\item{\emph{Distance Comparison}} is used to study how two or more time series differ at specific periods of time. It could be used in seasonality trends analysis or pattern extraction. 
For instance, in Meteorology, to compare the temperatures in Florida throughout 1940 and 2014, the high and low temperatures on each day of 1940 would be compared with each day of 2014.
\item{\emph{Modeling Training}} is to build a prediction model with the use of existing data. In modeling training, data are usually grouped into three parts: Training, Tests and Validation. For example, we can randomly select 10 years weather data to training a model and use the remained years' data for Tests and Validation. 
\item{\emph{Events Analysis}} is used to investigate the cause/effect of a special event. For instance, in telephone security, fraud can be detected by comparing the distributions of typical phone calls and of calls made from a stolen phone. 
\end{itemize}

\begin{figure}[!t]
\centering
\epsfig{file=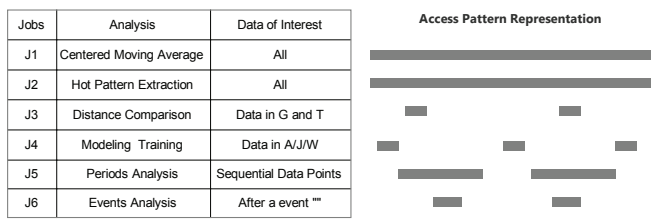, width=0.48\textwidth}
\caption{Common access patterns in selective bulk analysis.}
\label{fig:patterns}
\end{figure}

We illustrate the typical data access patterns for these data analysis methods in figure~\ref{fig:patterns}. In specific execution, all these methods could access parts or the overall data partitions. 

\vspace{0.02in}

\textbf{In-memory Big Data Processing}
Spark is a MapReduce framework for big data analysis. Spark introduces resilient distributed datasets (RDDs) to facilitate the programming of parallel applications. Each RDD represents a collection of data partitions that spreads across the cluster. 
We present a typical example for how a to run a spark application, which read a data from a file system, filter the error messages, and then we count its elements using map and reduce interface.

\vspace{-0.02in}
{\tt \small
\begin{verbatim}
   val file = spark.textFile("//data...")
   val errs = file.filter(_.contains("ERROR"))
   val ones = errs.map(_ => 1)
   val count = ones.reduce(_+_)
\end{verbatim}
}
\vspace{-0.02in}

\begin{figure}[h]
\centering
\epsfig{file=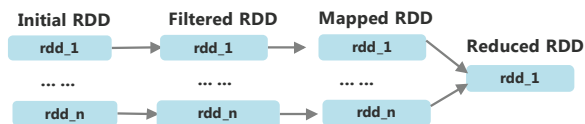, width=0.43\textwidth}
\caption{A example of data flow.}
\label{fig:rddflow}
\end{figure}

The data flow goes as Figure~\ref{fig:rddflow}. As we can see, a newly RDD will be formed with applying each data operation. To collect the lines including the text of \emph{error}, all the small data blocks (rdds) will call the filter operation and find the result. 

However, since selective bulk data analysis have its own characteristics and the default processing workflow in Spark can cause inefficiencies during execution. For instance, to perform a period analysis with the use of spark, a filter operation is usually needed to perform on all data partitions in order to prepare the specific period data. This requires us to scan and filter all the data partitions and costs extra memory to store the new generated data partitions. Such a filter operation is necessary because the content in each data partition is unknown to us without thorough scanning it. 


\section{In-Memory Content-aware Data Organization}
In order to efficiently support selective bulk data analysis in spark, 
we propose a novel content-aware method to 
 allow analysis programs to efficiently access their needed data without thorough scanning/filtering all the data partitions. 
There are two advantages for the content-aware method. (1) Memory efficiency: we don't need extra memory space to store the selective dataset, e.g. \emph{\_.filterRDD}
(2) Computation efficiency: data selection with content-aware method is much faster than the filter operation applied on all data partitions. 





\subsection{Table\_based Content\_Aware Data Organization} 
\begin{figure}[h]
\centering
\epsfig{file=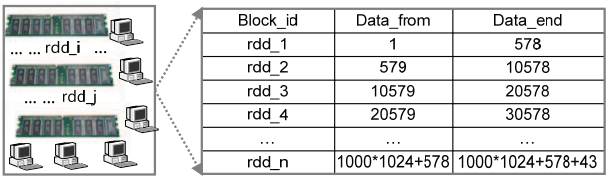, width=0.45\textwidth}
\caption{ Table-based content-aware data organization in selective bulk analysis.} 
\label{fig:table}
\end{figure}

To support selective bulk data processing efficiently, we record the metadata of each data block (rdd). In this paper, the metadata mainly refers to the data range, which is the major filter condition used in temporal/spatial data such as data with time property.   
An intuitive way to maintain the metadata for each data partition(block) is to use a table, similar to the technique adopted in database. The key and the value are the id of blocks and the data range of each block respectively, as shown in Figure~\ref{fig:table}. With the help of this metadata table, we can identify the specific data partitions given a range. For instance, if we need to find the data ranging from index $i$ to $j$, we can use a binary search to find which rdd contains the data item with index of $i$ and $j$ respectively, then all the rdds between them in the table are the targeted data items. 

However, such an intuitive method may encounter some challenges with the increasing number of data partitions (rdds). Firstly, the space complexity of table-based method is $O(m)$, where $m$ is the number of data partitions. This implies that the memory space will grow linearly with the increasing number of blocks. 
Secondly, the lookup time in table-based design is related to the size of table and the average lookup time should be $O(log$ $m)$. In reality, the metadata is usually used by application driver/scheduler which is usually based on a centralized architecture, the complexity of space and time should be as lighter as possible. 

\subsection{Compressed Index with Associated Search List (CIAS)}
As discussed, the table-based method could be inefficient regarding to the time and space issues. 
In this section, we propose a more efficient method to capture the mapping relationship between partitions' id and their data ranges. 
Our goal is to find a way such that the overhead on metadata organization and lookup does not increase with the size of real data or the data partitions.
We format our problem as, given the id of data partitions and their data ranges as shown in Figure~\ref{fig:table}, find a method to capture their relationship such that the memory cost is not affected by the size of table. 
Based on the relationship, we don't need to reside the table in memory and use binary search to lookup our target data.
 

To solve the problem, we propose a data structure called Compressed Index with Associated Search List (CIAS) to record the relationship between the data range with the id of data partitions. 
The design is based on the following facts, (1) the distributed bocks (rdds) in Spark usually have the same size, e.g, 32 MB or 64 MB.  
(2) data with time property such as time series have a fixed size on each periods, for instance, the weather data or stock prices.      
CIAS can represent the table in a compact way and lookup data range through a computation fashion. For instance, the table in Figure~\ref{fig:table} could be represent as following,   

\vspace{-0.02in}
{\tt \small
\begin{verbatim}
Compressed Index: 
578, 10000^1024, 43
AssociatedSearchList(ASL): 
578, 10240578, 10240621
\end{verbatim}
}
\vspace{-0.02in}

\section{Experiment}\label{exp}

We have conducted preliminary experiments on Marmot. 
\emph{Marmot} is a cluster of the PRObE on-site project~\cite{ref:probegarth} that is housed at CMU in Pittsburgh. The system has 128 nodes / 256 cores and each node in the cluster has dual 1.6GHz AMD Opteron processors, 16GB of memory, Gigabit Ethernet, and a 2TB Western Digital SATA disk drive. For our experiments, all nodes are connected to the same switch.
On \emph{Marmot}, Spark $[1.0.2]$ is installed as big data in-memory processing framework on all compute nodes running CentOS55-64 with kernel 2.6. 


\subsection{ Oseba Evaluation} 

\begin{figure}[h]
\centering
\epsfig{file=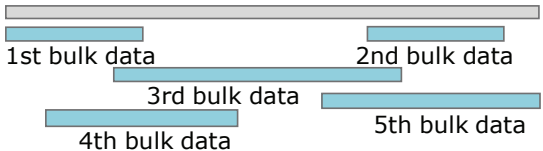, width=0.40\textwidth}
\caption{The pattern of five bulk data are selected during periods analysis.}
\label{fig:periods}
\end{figure}

To test Oseba, we use a benchmark application which interactively processes a data set on different periods. 
The experiments data is a time series, which has the similar data format to the climate data, e.g, \emph{time, temperature, humidity, wind speed and direction}.  
The size of our dataset is around 480 MB and the data is partitioned into 15 partitions after loading into memory. We process the dataset via two methods. The first method is to use the default data processing interface and method in Spark, in which we firstly load/reside the data into memory, then we apply \emph{filter} operation to obtain our target period data and finally we perform statistic analysis on the selective bulk data. 
The second method is our proposed method: Oseba, which records the content range for each data partition. Instead of scanning all data partitions during filter operation, we can find the data partition which contain our target data with the use of Oseba. In our experiment, 5 bulk data from different periods are selected to do analysis, as shown in Figure~\ref{fig:periods}. For each period, we do three basic statistic analysis on \emph{temperature} property: computing the \emph{max, mean and standard deviation} of the selected elements. We mainly compare the performance of memory and processing time in this paper. 

\begin{figure}[!t]
\centering
\epsfig{file=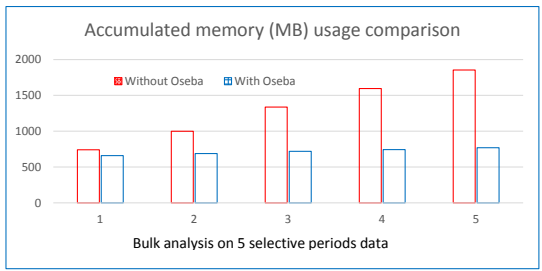, width=0.40\textwidth}
\caption{The memory cost comparison for selective bulk analysis on five periods.}
\label{fig:memory}
\end{figure}

The execution includes five phases according to the selected five periods. After finishing each phase, we monitor the total used memory.  
The memory comparison is shown in Figure~\ref{fig:memory}. With the use of default method, we can find that the memory usage is increasing. This is because after each phase, more RDDs are created and they are resident in memory by default. The final accumulated used memory is around 1800MB, which is about 3.8X to the raw input data.
On the other hand, with the use of Oseba, we can achieve a much lower memory cost than the default method. The used memory is almost not increasing. This is because we don't need to save "filter" RDDs in comparison to the default method. In general, we can find the memory cost is half that of without Oseba after the analysis on the third period, and a third for the fifth period. This shows our method is efficient during bulk data analysis.  

\begin{figure}[!t]
\centering
\epsfig{file=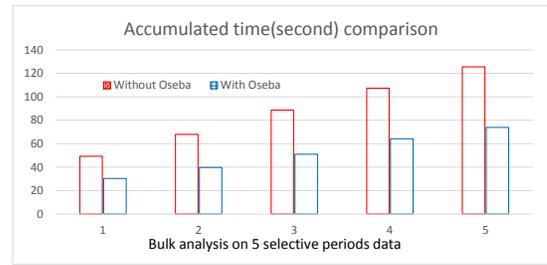, width=0.40\textwidth}
\caption{The time cost comparison for selective bulk analysis on five periods.}
\label{fig:time}
\end{figure}

We also collected the accumulated time based on the five phases and the result is shown in Figure~\ref{fig:time}.
Clearly, we can find less time is cost with the use of Oseba in comparison to that of without Oseba. There is a little improvement for the first analysis. After that, the processing time gap become much bigger. The total processing time is more than 120 seconds without the use of Oseba while that is around 70 second with the use of Oseba.   
This could be explained that the thorough scanning during \emph{filter} operation is expensive for selective bulk data analysis. In fact, a larger size of raw data can result in a bigger time consumption during selecting bulk data.     

\section{Related work}\label{rel}

There are a great deal of frameworks or systems  that are proposed to manage and process big data~\cite{dean2008mapreduce, isard2007dryad, tran_dataNet, power2010piccolo,zhang2016sapprox}. 
The Hadoop Distributed File System (HDFS) is an open source community response to the Google File System (GFS), specifically for the use of MapReduce style workloads~\cite{dean2008mapreduce}.
Dryad~\cite{isard2007dryad} 
and Spark 
are two other frameworks to support big data processing with the similar interface to MapReduce. Spark can allow data to be repeatedly and interactively processed in distributed memory. 
and Sparkler~\cite{Sparkler} extends Spark to support distributed stochastic gradient descent. 
However, these systems usually apply operations to all data items, which is inconvenient for selective bulk data analysis since only a subset of data is involved. 
Also, Pregel~\cite{malewicz2010pregel,zhang2015achieving} supports iterative graph applications and HaLoop are iterative MapReduce runtimes. 
Moreover, there are systems that can support fine-grained data processing. Example of these systems are keyvalue stores~\cite{RAMClouds}, databases, and Piccolo~\cite{power2010piccolo} and they provide interfaces to support fine-grained data items/cells updates/processing.
However, these frameworks and systems need extra cost for maintaining reliability as discussed in Spark, while bulk data analysis are more about coarse-grained level data processing.

\section{Conclusion}
\label{con}

In this paper, we investigate the problems of selective bulk analysis on in-memory big data processing frameworks, e.g, Spark. Due to the missing information of blocks' content, 
selective bulk analysis programs needs to scan thorough the whole data partitions in order to find the valid  partitions, resulting in extra computation and memory overheads. To address this problem, we propose a content-aware data organization method to help selective bulk analysis. With the use of our method (Oseba), selective bulk data analysis program can easily identify their needed data. 
We conduct some preliminary experiments for our proposed method on PRObE’s Marmot and the experimental results show the promising performance of Oseba. 

\section*{ Acknowledgments}

This material is based upon work supported by the National Science Foundation under the following NSF program: Parallel Reconfigurable Observational Environment for Data Intensive Super-Computing and High Performance Computing (PRObE).

\bibliographystyle{abbrv}
\bibliography{Opass}

\begin{thebibliography}{10}

\bibitem{cite:PatternR}
C.~M. Bishop.
\newblock {\em Pattern Recognition and Machine Learning (Information Science
  and Statistics)}.
\newblock Springer-Verlag New York, Inc., Secaucus, NJ, USA, 2006.

\bibitem{dean2008mapreduce}
J.~Dean and S.~Ghemawat.
\newblock Mapreduce: simplified data processing on large clusters.
\newblock {\em Communications of the ACM}, 51(1):107--113, 2008.

\bibitem{ref:probegarth}
G.~Gibson, G.~Grider, A.~Jacobson, and W.~Lloyd.
\newblock Probe: A thousand-node experimental cluster for computer systems
  research.

\bibitem{isard2007dryad}
M.~Isard, M.~Budiu, Y.~Yu, A.~Birrell, and D.~Fetterly.
\newblock Dryad: distributed data-parallel programs from sequential building
  blocks.
\newblock {\em ACM SIGOPS Operating Systems Review}, 41(3):59--72, 2007.

\bibitem{Sparkler}
B.~Li, S.~Tata, and Y.~Sismanis.
\newblock Sparkler: Supporting large-scale matrix factorization.
\newblock In {\em Proceedings of the 16th International Conference on Extending
  Database Technology}, EDBT '13, pages 625--636, New York, NY, USA, 2013. ACM.

\bibitem{malewicz2010pregel}
G.~Malewicz, M.~H. Austern, A.~J. Bik, J.~C. Dehnert, I.~Horn, N.~Leiser, and
  G.~Czajkowski.
\newblock Pregel: a system for large-scale graph processing.
\newblock In {\em Proceedings of the 2010 ACM SIGMOD International Conference
  on Management of data}, pages 135--146. ACM, 2010.

\bibitem{RAMClouds}
J.~Ousterhout, P.~Agrawal, D.~Erickson, C.~Kozyrakis, J.~Leverich,
  D.~Mazi{\`e}res, S.~Mitra, A.~Narayanan, G.~Parulkar, M.~Rosenblum, et~al.
\newblock The case for ramclouds: scalable high-performance storage entirely in
  dram.
\newblock {\em ACM SIGOPS Operating Systems Review}, 43(4):92--105, 2010.

\bibitem{power2010piccolo}
R.~Power and J.~Li.
\newblock Piccolo: Building fast, distributed programs with partitioned tables.
\newblock In {\em OSDI}, volume~10, pages 1--14, 2010.

\bibitem{price2000comparison}
D.~T. Price, D.~W. McKenney, I.~A. Nalder, M.~F. Hutchinson, and J.~L.
  Kesteven.
\newblock A comparison of two statistical methods for spatial interpolation of
  canadian monthly mean climate data.
\newblock {\em Agricultural and Forest meteorology}, 101(2):81--94, 2000.

\bibitem{Wang2014}
J.~Wang, P.~Shang, and J.~Yin.
\newblock Draw: A new data-grouping-aware data placement scheme for data
  intensive applications with interest locality.
\newblock In {\em Cloud Computing for Data-Intensive Applications}, pages
  149--174. Springer New York, 2014.

\bibitem{tran_dataNet}
J.~Wang, X.~Zhang, J.~Yin, H.~Wu, and D.~Han.
\newblock Speed up big data analytics by unveiling the storage distribution of
  sub-datasets.
\newblock {\em IEEE Transactions on Big Data}, 2016.

\bibitem{yin_time}
J.~Yin, Y.-W. Si, and Z.~Gong.
\newblock Financial time series segmentation based on turning points.
\newblock In {\em System Science and Engineering (ICSSE), 2011 International
  Conference on}, pages 394--399. IEEE, 2011.

\bibitem{zhang2016sapprox}
X.~Zhang, J.~Wang, and J.~Yin.
\newblock Sapprox: enabling efficient and accurate approximations on
  sub-datasets with distribution-aware online sampling.
\newblock {\em Proceedings of the VLDB Endowment}, 10(3):109--120, 2016.

\bibitem{zhang2015achieving}
X.~Zhang, R.~Wang, X.~Chen, J.~Wang, T.~Lukasiewicz, and D.~Han.
\newblock Achieving up to zero communication delay in bsp-based graph
  processing via vertex categorization.
\newblock In {\em Networking, Architecture and Storage (NAS), 2015 IEEE
  International Conference on}, pages 112--121. IEEE, 2015.

\end{thebibliography}
\balance

\end{document}